\documentclass{aa}
\usepackage{graphicx,times}

\begin{document}

\title{Cas--A}
\title{The mass and energy budget of Cassiopeia A}

\author{R.Willingale\inst{1}, J.A.M.Bleeker\inst{2},
K.J. van der Heyden\inst{2}, J.S.Kaastra\inst{2}}

\offprints{}

\institute{Department of Physics and Astronomy, University of Leicester, 
University Road, Leicester LE1 7RH, UK
\and
SRON National Institute for Space Research, Sorbonnelaan 2,
3584 CA Utrecht, The Netherlands
\\email: rw@star.le.ac.uk ; J.A.M.Bleeker@sron.nl ; K.J.van.der.Heyden@sron.nl
; j.s.kaastra@sron.nl\\}

\titlerunning{The mass and energy budget of Cas A}
\authorrunning {R.Willingale et al.}

\date{Received ; accepted }

\abstract{
Further analysis of X-ray spectroscopy results
(Willingale et al. 2002) recently obtained from the MOS CCD
cameras on-board XMM-Newton provides a detailed description of the hot and cool
X-ray emitting plasma in Cas A.
Measurement of the Doppler broadening of the X-ray emission lines
is consistent with the expected ion velocities,
$\sim1500$ km\,s$^{-1}$ along the line of sight, in the post shock plasma.
Assuming a distance of 3.4 kpc,
a constant total pressure throughout the remnant
and combining the X-ray observations with optical measurements
we estimate the total remnant mass as 10 $M_{\sun}$ and the
total thermal energy as $7\times10^{43}$ J. We derive the differential mass 
distribution as a function of ionisation age for the hot and cool
X-ray emitting components. This distribution is consistent with a hot component 
dominated by swept up mass heated by the primary shock and a cool
component which are ablated clumpy ejecta material which were and 
are still being 
heated by interaction with the preheated swept up material.
We calculate a balanced mass and
energy budget for the supernova explosion giving a grand total of
$1.0\times10^{44}$ J in
an ejected mass; approximately $\sim0.4$ $M_{\sun}$ of the ejecta were
diffuse with an initial rms velocity $\sim1.5{\times}10^{4}$~km\,s$^{-1}$
while the remaining $\sim1.8$ $M_{\sun}$ were clumpy with an initial
rms velocity of $\sim2400$~km\,s$^{-1}$.
Using the Doppler velocity measurements of the X-ray spectral lines
we can project the mass into spherical coordinates about the remnant. This
provides quantitative evidence for mass and energy
beaming in the supernova explosion.
The mass and energy occupy less than 4.5 sr ($<$40\% of the available
solid angle) around the remnant and 64\% of the mass occurs in two
{\em jets} within 45 degrees of a jet axis. We calculate a swept up mass 
of 7.9 $M_{\sun}$ in the emitting plasma
and estimate that the total mass lost from the progenitor
prior to the explosion could be as high as $\sim20$ $M_{\sun}$. We suggest 
that the progenitor was a Wolf-Rayet star that formed a 
dense nebular shell before the supernova explosion. 
This shell underwent heating by the primary shock which was
energized by the fast diffuse ejecta.}
\maketitle

\section{Introduction}

If we can measure the total mass, the temperature and
the bulk velocity of material in a young SNR we can estimate the total
energy released by the SN explosion. Coupling this with
Doppler measurements we
can deproject the mass and energy from the plane of the sky into
an angular distribution around the centre of the SN.
Here we present further analysis of XMM-Newton data (Willingale et al.
2002) that provides a quantitative assessment
of the mass and energy distribution around Cas A.
There is a growing body of evidence that the core collapse of
massive stars is an asymmetric process.
Spectra of supernovae are polarized, neutron stars produced
in supernovae have high velocities, mixing of high-Z radioactive
material from the core with hydrogen-rich outer layer of ejecta
is very rapid, high velocity bullets have been observed in the Vela
SNR (Aschenbach et al. 1995)
and Cas A itself (Markert et al. 1983, Willingale et al. 2002)
is composed of two oppositely directed jets.
The analysis presented here confirms the non-spherical
nature of the Cas A SNR and also provides
details about the ionization state of the X-ray emitting plasma
and the total energy and mass budget of the SN explosion.

\section{Composition and dynamics of the plasma}

The spectral fit data from Willingale et al. (2002)
provide electron
temperature $kT_{\rm e}$, emission integral $E_{I}=\int n_{\rm e}n_{\rm H}dv$,
ionization age $I_{\rm t}=\int n_{\rm e}dt$ and elemental abundances
for two plasma components (hot and cool) over a grid of $20\times 20$
arc second pixels covering the face of Cas A. If $n_{\rm e}$ is the electron
density and $n_{\rm H}$ is the hydrogen density,
using the elemental abundances
and assuming a fully ionised plasma we can calculate the number of electrons
per hydrogen atom $R_{\rm e}=n_{\rm e}/n_{\rm H}$, the effective number of
protons and neutrons
(baryon mass) per hydrogen atom $R_{\rm m}=n_{\rm m}/n_{\rm H}$
and the number of baryons per hydrogen atom $R_{\rm i}=n_{\rm i}/n_{\rm H}$.
Table \ref{tab1}
summarises these plasma parameters. The mean and rms values were calculated by
weighting with the shell volume associated with
each pixel (see below).
\begin{table}[!htb]
\caption[]{Mean and rms values for those plasma parameters which are not
explicitly dependent on the plasma volume, across the face of the remnant.
The values are weighted by the shell volume associated with each pixel.
$R_{\rm m}/R_{\rm i}$ is the mean mass per baryon in units of proton mass.}
\centering
\begin{tabular}{|c|cccc|}
\hline
 & $R_{\rm e}$ & $R_{\rm m}$ & $R_{\rm i}$ & $R_{\rm m}/R_{\rm i}$\\
\hline
hot    & $1.23\pm0.02$  & $1.47\pm0.05$ & $1.10\pm0.03$ & $1.33\pm0.04$ \\
cool   & $289\pm222$ & $582\pm449$ & $37\pm27$ & $15.6\pm0.7$ \\
\hline
\end{tabular}
\label{tab1}
\end{table}
The cool plasma component
used for the spectral modelling was assumed to be
oxygen rich rather than hydrogen
rich with all the elemental abundances for the elements heavier than
Helium being multiplied by a factor of 10000. Therefore the electron
density and baryon mass per hydrogen atom are high for this
component and because of the variations in abundance of the heavy elements
there is considerable scatter in $R_{\rm e}$, $R_{\rm m}$ and $R_{\rm i}$.

Using the combination of
measured Doppler shifts and sky positions for each pixel
we were able to estimate the radial distribution of emissivity within
the spherical cavity surrounding Cas A (Willingale et al. 2002).
The bulk of the emission 
is confined to a spherical shell of radius 60 to 170 arc seconds.
Assuming a distance of 3.4 kpc
we can calculate the emission volume within the spherical shell
associated with
each pixel. We know from the high resolution Chandra image
of the remnant, Hughes et al. (2000), that
the X-ray emission is broken into tight knots and that the plasma
doesn't fill the spherical shell.
We have therefore assumed
filling factors for the components $\eta_{\rm hot}$ and
$\eta_{\rm cool}$.
Adopting
mean values for $R_{\rm e}$, $R_{\rm m}$ and $R_{\rm i}$ over the plasma volume 
and as a function of
time we can estimate the electron density ($n_{\rm e}$), the hydrogen density 
($n_{\rm H}$), the effective ionisation age ($t$),
the total emitting mass ($M$) 
the thermal pressure ($P_{\rm th}$) and total thermal energy ($E_{\rm th}$) by 
using:
\begin{equation}
n_{\rm e}=\sqrt{E_{\rm I}R_{\rm e}/(V\eta)} \label{eq1}
\end{equation}
\begin{equation}
n_{\rm H}=\sqrt{E_{\rm I}/(V\eta R_{\rm e})} \label{eq2}
\end{equation}
\begin{equation}
t=I_{\rm t}\sqrt{V\eta/(E_{\rm I}R_{\rm e})} \label{eq3}
\end{equation}
\begin{equation}
M=m_{\rm p}R_{\rm m}E_{\rm I}t /I_{\rm t}=m_{\rm p}R_{\rm m}\sqrt{E_{\rm 
I}V\eta/R_{\rm e}} 
\label{eq4}
\end{equation}
\begin{equation}
P_{\rm th}=k(T_{\rm i}R_{\rm i}+T_{\rm e}R_{\rm e})\sqrt{E_{\rm I}/(V\eta R_{\rm 
e})} 
\label{eq5}
\end{equation}
\begin{equation}
E_{\rm th}=(3/2)k(T_{\rm i}R_{\rm i}+T_{\rm e}R_{\rm e})\sqrt{E_{\rm 
I}V\eta/R_{\rm 
e}} 
\label{eq6}
\end{equation}
Here $m_{\rm p}$ is the proton mass, $T_{\rm i}$ is the ion temperature,
$V$ is the total plasma volume and
$\eta$ is a filling factor within that volume.

\begin{figure}[!htb]
\centering
\includegraphics{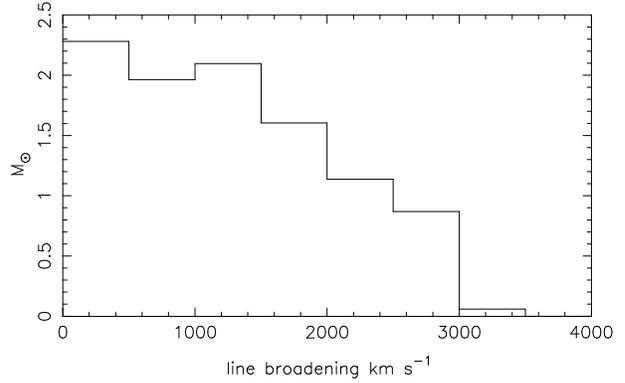}
\caption{Mass distribution of line broadening velocity derived from
$\sim195$ spectral fits across the face of Cas A.}
\label{fig1}
\end{figure}

In the spectral fitting we also included
a Doppler broadening term to fit the line profiles. Fig. \ref{fig1}
shows the mass distribution of the fitted line broadening
velocity derived using the mass estimates described below.
The rms velocity of the distribution
is $1490\pm110$ km\,s$^{-1}$. This represents the rms velocity
broadening measured from $\sim195$ spectral fits over the face of the
remnant. Some of the
line broadening may not be due to Doppler
but could be introduced by variations in the line blending
as a function of temperature which were not accurately modelled
using just two temperature components.
By looking at the change in line blends over
the temperature range of the spectral fits
we estimate this introduces a systematic rms error of
$\le440$ km\,s$^{-1}$. Estimation of broadening of the line profiles
also depends on accurate modelling of the spectral response of the
MOS detectors. This is known to an accuracy of a few eV which introduces
a possible systematic error of $\pm 500$ km\,s$^{-1}$.
These systematic errors are small compared to
statistical error on the individual spectral fits and
the distribution of velocities shown in Fig.~\ref{fig1} is
dominated by Doppler shift due to the motion of ions in the plasma.
The width of the distribution is due to the large spread of ion 
temperatures (velocities) within the remnant volume.
The spectral fitting gives us a direct measurement of the electron
temperature $T_{\rm e}$ in the plasma but not the ion temperature $T_{\rm i}$.
Laming (2001) provides predictions of the $T_{\rm i}$ and $T_{\rm e}$ for the
forward and reverse shocks in Cas A as a function of shock time
after the explosion. Using typical ages of the hot and cool 
components derived below (Table \ref{tab3}) and assuming the hot
component is characteristic of the forward shock and
the cool component is characteristic of the reverse shock we estimate
$(T_{\rm i}/T_{\rm e})_{\rm cool}=135$ and $(T_{\rm i}/T_{\rm e})_{\rm 
hot}=10.5$.
These ratios are not very sensitive to the ages assumed.
The rms thermal velocity along the line of sight
is given by $v_{\rm th}=\sqrt(kT_{\rm i}/m_{\rm b})$ where
$m_{\rm b}$ is the mean mass of baryons in the plasma. Therefore using
the measured electron temperature, the mean mass per baryon listed
in Table \ref{tab1} and $T_{\rm i}/T_{\rm e}$ ratios predicted by Laming we
can estimate the ion velocity for comparison with the measured
Doppler velocities.

In Willingale et al. (2002) the Doppler
shifts of the prominent emission lines in the X-ray spectrum were used to
derive a linear approximation to the radial plasma velocity
within the remnant volume,
\begin{equation}
v(r)=\frac{v_{\rm s}}{(r_{\rm s}-r_{\rm o})}(r-r_{\rm o})
\label{eq7}
\end{equation}
where the shock radius $r_{s}=153$\arcsec, the velocity falls to
zero at $r_{\rm o}=53$\arcsec\ and the velocity of the plasma just behind the
shock is $v_{\rm s}=2600$ km\,s$^{-1}$. We can use this relationship to estimate
the rms radial velocity $v_{\rm r}$ of the plasma components.

Table~\ref{tab2} gives a summary of the ion velocity results.
\begin{table}[!htb]
\caption[]{The measured electron temperature and
ion velocity components due to thermal motion (along the line of sight, Doppler 
broadening) and Doppler shifts of the spectral lines. The velocities are rms 
mass-weighted values
using the mass associated with each pixel.}
\centering
\begin{tabular}{|c|ccc|}
\hline
 & $T_{\rm e}$ & $v_{\rm th}$ & $v_{\rm r}$ \\
 &   keV & km\,s$^{-1}$  & km\,s$^{-1}$ \\
\hline
hot  & $3.27\pm0.86$ & 1575 & 1740  \\
cool & $0.45\pm0.10$ &  608 & 1780  \\
\hline
\end{tabular}
\label{tab2}
\end{table}
The electron temperatures measured are considerably lower than Laming's
predictions especially for the cool component. This may be because the
modelling assumes uniform density without clumping.
The thermal velocity
for the cool plasma is surprisingly low because the mean mass
per baryon is rather large for this component (very close to
pure oxygen as assumed in the Laming 2001 predictions).
There is reasonable agreement between the predicted thermal ion velocity
and the ion velocity measured from Doppler broadening of the lines
indicating that the predicted $T_{\rm i}/T_{\rm e}$ ratios are about right.
What we actually measure in the spectral fitting is the weighted average
of the Doppler broadening from the hot and cool line components
combined. This
is predicted to be $\sim1460$ km\,s$^{-1}$ compared with the measured value of 
$1490\pm110$ km\,s$^{-1}$.
However, some of the Doppler broadening could be due to chaotic motion at
large scales rather than microsopic thermal motion of the ions.
If this were the case we would require lower $T_{\rm i}$ values.
Chaotic proper motions, over and above linear expansion,
have been observed for radio knots, Anderson \&
Rundick (1995), (see Fig. 6 ibid) but they are small compared with
the radial component. Fortunately, when calculating the energy
associated with these velocities (see below) it doesn't matter whether they
are attributable to thermal motion or turbulence.

\section{Mass and energy of the plasma}

In order to calculate the mass associated with the hot and cool components we
must estimate the filling factors. We can do this if
we assume that the total pressure is the same in each of the pixels across the
face of the remnant and that the cool and hot phases are in pressure
equilibrium. The pressure in the plasma has three components, thermal, ram and
magnetic. We don't know the magnetic condition of the hot and cool components 
but it is reasonable to assume that the magnetic
pressure is proportional to the 
thermal pressure.
We were unable to detect a large systematic difference between 
the radial velocities of the two components (see Table~\ref{tab2}) and the 
turbulent velocities are probably small compared with the thermal velocities so 
the turbulent ram pressure is not important. The ram pressure due to the 
bulk motion should be comparable to the thermal pressure since $v_{\rm 
r}\approx v_{\rm th}$ (see Table~\ref{tab2}). In calculating the pressure we 
should use $P = P_{\rm th} + P_{\rm R} + P_{\rm mag}$, where $P_{\rm th}$
is the 
thermal pressure given by eq.~(\ref{eq5}),
$P_{\rm R}$ is the total ram pressure 
and $P_{\rm mag}$ is the magnetic pressure. If we 
restrict $P = P_{\rm th}$, we find that a minimum 
pressure of $7.91\times10^{-8}$ Pa (N m$^{-2}$) gives the maximum possible 
filling factor of $\eta_{\rm hot}+\eta_{\rm cool}\approx 1.0$ peaking near the 
Western limb of the remnant. Using the minimum pressure equilibrium filling 
factors we have derived values for the electron and baryon densities,
ionisation age, total emitting mass and total thermal energy given in Table 
\ref{tab3}. Including the ram pressure due to the bulk motion changes $P$ by a 
factor 2. This does not influence the results  since the total pressure would 
scale to allow for a maximum $\eta \approx 1$.
If we allow the maximum filling factor to drop below 1.0 or we
assume a different distance to the remnant the values and ranges in Table
\ref{tab3} will, of course, change.

\begin{table}[!htb]
\caption[]{Estimates of the volume filling factors,
mean electron and baryon densities,
ionisation age, total emitting mass and total thermal energy
for the hot and cool components
within the spherical shell radius 60-170\arcsec.
The baryon density $n_{\rm m}$ is expressed in units of proton mass. The mean 
and $\pm$ values given are shell volume-weighted.
The ionization ages are the median values from the
mass ionisation time distribution (see Fig. \ref{fig3}).
The ranges quoted for these ages
are the 25 and 75 percentiles of the same distribution.
The ranges given for the mass and thermal energy are discussed in the text.}

\centering
\begin{tabular}{|c|cccccc|}
\hline
 & $\eta$ & $n_{\rm e}$ & $n_{\rm m}$ & $t$ & $M_{\sun}$ & $E_{\rm th}$  \\
 &        & cm$^{-3}$& cm$^{-3}$& yrs &           &  $10^{43}$ J \\
\hline
h  & 0.31 & 16 & 19 & 131 &  8.31 & 6.82 \\
     & $\pm0.20$ &$\pm3$  & $\pm4$ & 101-182  & 7.42-9.20 & 6.09-7.55 \\
c & 0.009 & 61 & 123 & 80 & 1.70 & 0.20 \\
   & $\pm0.014$  & $\pm15$  &$\pm30$  & 20-273  & 1.60-1.80 & 0.19-0.21  \\
\hline
\end{tabular}
\label{tab3}
\end{table}

We repeated the above analysis using constant values for the
filling factors ($\eta_{\rm hot}=0.31$, $\eta_{\rm cool}=0.009$) instead of
assuming pressure equilibrium. The results were very
similar to those in Table \ref{tab3}.
The critical factors that effect the results are the
the magnitude of the pressure (or filling factor) and the implied large
difference between the filling factors for the hot and cool components.
Actually the results don't require strict pressure equilibrium between the
hot and cool components. We simply require
the same mean pressure in the two components for a given pixel.
The results are also dependent on the volume of the emitting shell. This
could be as small as 80-170 arc seconds or as large as 50-180 arc seconds.
If we change the shell parameters the pressure
must change in order to give a maximum filling factor of $\sim1.0$. Changing
the shell volume within the allowed limits has exactly the same effect
as changing the pressure. The corresponding pressure range is
$(7.48-8.34)\times10^{-8}$ Pa. Using this range we can estimate
ranges on the total mass and energy as shown in Table \ref{tab3}.

\begin{figure}[!htb]
\centering
\includegraphics[width=8cm]{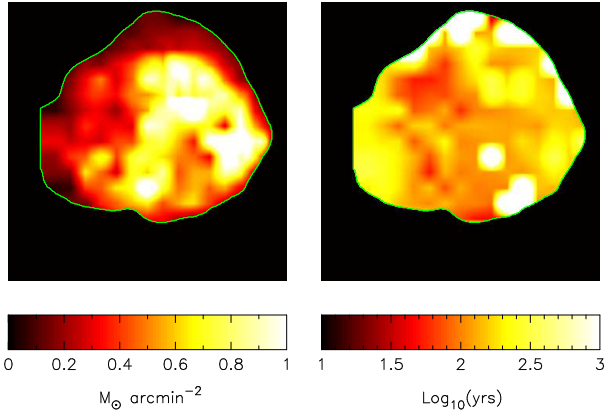}
\caption{The maps of mass and ionization age of the hot component.
The outer contour line indicates the extent of sky coverage with
good statistics for spectral fitting.}
\label{fig2}
\end{figure}

\begin{figure}[!htb]
\centering
\includegraphics[width=8cm]{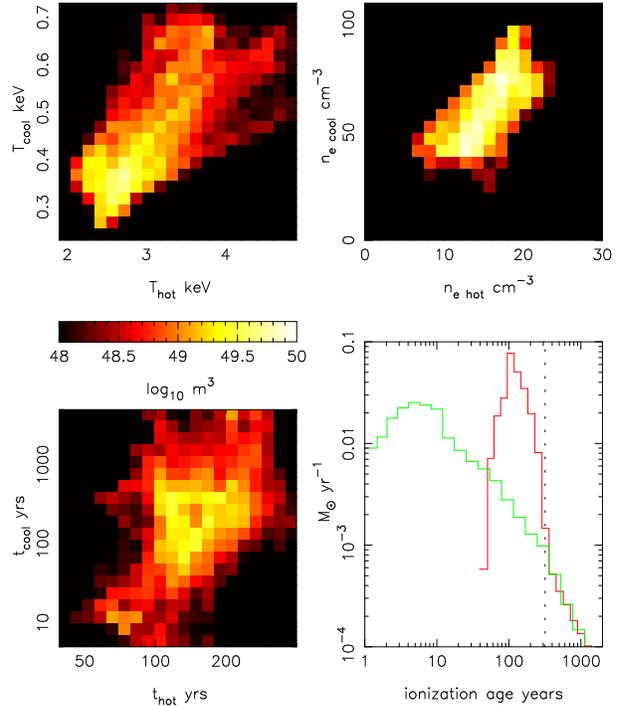}
\caption{The distribution of shell volume in the temperature,
electron density and ionisation age planes. Note that the volume
is plotted on a logarithmic scale to reveal the regions with very
low volume. The ionisation age scales are logarithmic showing the
spread of the volume over several orders of magnitude of the
ionisation time. The bottom right panel shows the differential
mass distribution of ionization ages.
The hot component is shown in red and the cool component is shown in green.
The vertical dotted line indicates the age of the remnant (320 years).
\label{fig3}}
\end{figure}

The maps of the mass distribution and ionization age of the hot component
are shown in Fig. \ref{fig2}.
The larger ages tend to lie around the perimeter while the ages across the
central region are relatively
constant with a minimum of $\sim45$ years.
Fig. \ref{fig3} shows the distribution of the shell volume in the
temperature ($T_{\rm hot}:T_{\rm cool}$), electron density ($n_{\rm e 
hot}:n_{\rm e cool}$)
and ionisation age ($t_{\rm hot}:t_{\rm cool}$) planes. The electron density
and to some extent the temperatures are correlated over the shell volume.
However the ionisation ages are not. Note that the ionisation ages
are plotted on logarithmic axes. The spread of age is much larger for
the cool component than the hot.
The differential
mass distribution as a function of ionization age for the two components
is also shown in Fig. \ref{fig3}. In this distribution
we do see a marked difference between the two plasma components.
The hot 
component shows a relatively sharp peak at an ionization age of $\sim100$ 
years, whereas the cool component has a broader distribution with a median age 
of $\sim80$ 
years. These profiles indicate that the hot plasma was shock heated over a 
century ago and the heating process is already complete. The cool plasma, 
however, has been shocked more recently and the heating process is still
ongoing. A small fraction of the mass in both components has an ionization
age greater than the true age of the remnant, 320 years,
because we have assumed that the densities are constant with time.
This is clearly not the case especially when we extrapolate back to 
the early stages of the remnant expansion.

\section{Projection about the SNR centre}

\begin{figure}[!htb]
\centering
\includegraphics[width=5cm]{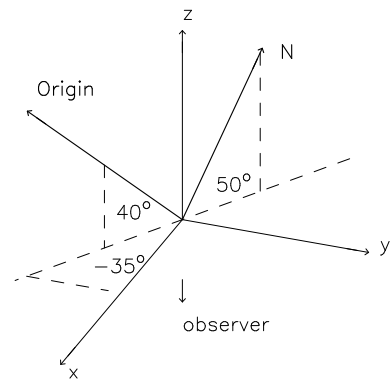}
\caption{The {\em axial} spherical coordinate system with respect
to the plane of the sky, $x$ East and $y$ North, and the line
of sight $z$.}
\label{fig4}
\end{figure}

\begin{figure}[!htb]
\centering
\includegraphics{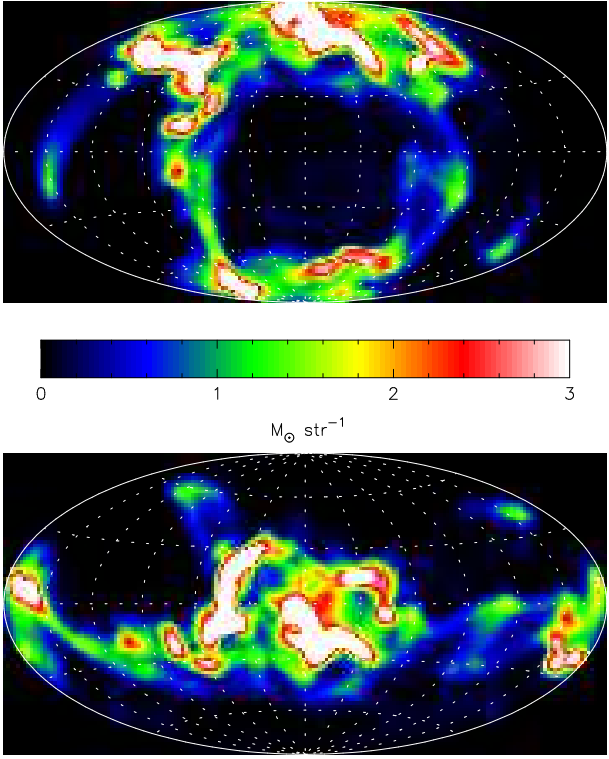}
\caption{The angular distribution of mass surrounding the centre
of Cas A - Aitoff equal area projection. Upper panel in the {\em axial}
projection and lower panel in the {\em equatorial} projection.}
\label{fig5}
\end{figure}

Using the pixel positions with respect to the centre of the remnant
on the plane of the sky and the line-of-sight positions provided
by the Doppler measurements we can determine the angular distribution
of the mass about the centre of the remnant. With $x$ East, $y$ North
and $z$ pointing away from the observer we define an {\em axial} spherical
coordinate system as shown in Fig. \ref{fig4}. The polar axis is labelled
$N$ and points towards the receding mass in the North of the 
remnant. The origin on the equator is in the South East quadrant
away from the observer.
The upper panel of
Fig. \ref{fig5} shows the mass distribution projected in this {\em axial}
coordinate system in Aitoff projection.
It is clear that the entire angular distribution
of the emitting mass lies in a band around the remnant with
enhancements at the poles in the North and South.
(as has been suggested by many observations in the past, see for example
Markert et al. 1983).
The band of mass is relatively weak when it crosses the equator and
there are large solid angle areas around the origin and
the anti-centre on the equator where there is very little mass. 

\begin{figure}[!htb]
\centering
\includegraphics[width=8cm]{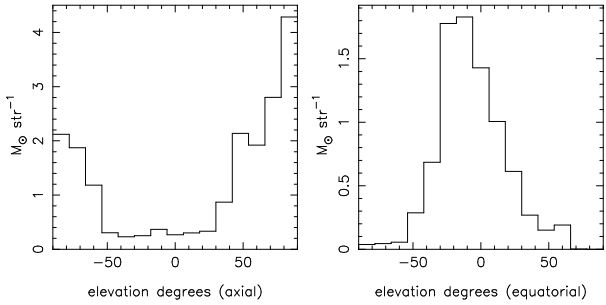}
\caption{The mass per steradian as a function of elevation angle.
Left-hand panel in the {\em axial} projection, South -ve elevation,
North +ve elevation. Right-hand panel in the {\em equatorial}
projection.}
\label{fig6}
\end{figure}

The total mass in the Southern hemisphere is $3.3$ $M_{\sun}$
and in the Northern hemisphere is $6.7$ $M_{\sun}$
so the split between the two hemispheres is not equal.
Half the mass in the South is contained in $0.63$ steradians, a sky fraction
of 0.10 and 90\% is contained in $2.0$ steradians (fraction 0.32).
Half the mass in the North is contained in $0.75$ steradians, a fraction
of 0.12 and 90\% is contained in $2.5$ steradians (0.39).
The left-hand
panel of Fig. \ref{fig6} shows the mass per steradian as a function of elevation
angle in the {\em axial} coordinate system.
64\% of the mass is contained within a double cone
of half angle 45\degr and the mass density
peaks at the poles.

The mass distribution was reprojected into an {\em equatorial}
spherical coordinate
system in which the equator lies around the band of mass seen in the top
panel of Fig. \ref{fig5}. The North pole was 
shifted to lie at the origin on the equator.
The lower panel in Fig. \ref{fig5} shows this new projection.
The right-hand panel of
Fig. \ref{fig6} shows the distribution about the equatorial plane.
In this reprojection a fraction of 0.85 of the mass is confined to 
within $\pm30$\degr of the equatorial plane.

The concentration of mass into a small fraction of the available
solid angle as shown in Figs. \ref{fig5} and \ref{fig6}
is consistent with the mean filling factor of $\eta_{\rm hot}=0.31$
(see Table \ref{tab3})
derived from the minumum equilibrium pressure corresponding
to a maximum filling factor of $\approx1.0$.
Pressure equilibrium dictates that the filling factor for the
cool component is very low, $\eta_{\rm cool}=0.009$.
This is consistent with the
original rationale for the spectral modelling in which the hot
component is assumed to be dominated by surrounding medium heated
by the primary shock and potentially enriched by diffuse ejecta, whereas the 
cool component is assumed to
be clumpy ejecta heated by the reverse shock. The differential mass
distributions with ionisation age are also consistent with this picture.
The bulk of the hot component forms a peak with
an ionisaton age in the range 100-180
years while the cool component has a much broader distribution
stretching back
to $<10$ years, probably indicative of a heating process which
is still in progress.

\section{The mass and energy budget\label{sect5}}

The maximum total X-ray emitting mass consistent with the data assuming
pressure equilibrium within the remnant volume is $10.0\pm0.7$ $M_{\sun}$.
The maximum total thermal energy visible is $7.0\pm0.5\times10^{43}$ J which
is a sizeable fraction of the typical total energy released from a
SN explosion. The remainder of the energy in the remnant is kinetic.
We can estimate
this kinetic energy using the expansion velocities discussed above.
To estimate the total
mass and kinetic energy in the optical FMKs we have assumed a generous
hydrogen density of $10^{3}$ cm$^{-3}$, (probably reasonably consistent with
pressure equilibrium)
a total of 120 knots, a knot size of 2\arcsec\ and a velocity of
5290 km\,s$^{-1}$
(see Reed et al. 1995 and Anderson and Rudnick 1995). Since we have
no information about chaotic velocities of optical knots we have
assumed these to be zero.
Table \ref{tab4} summarises the resulting mass and energy budget for
the remnant.
\begin{table}[!htb]
\caption[]{The mass and energy budget for the remnant. $E_{\rm th}$ is
the thermal energy and $E_{r}$ is the radial expansion kinetic energy. 
The hot Ejecta entry (2) are the heavy elements in the hot component, while the 
hot CSM entry (1) is the fraction of the hot component which
is not diffuse ejecta
and is presumed to have been circumstellar material at the time of the
explosion. The warm ejecta (3) are from the cool X-ray component and the cold 
ejecta (4) are the optical component. The hot diffuse entry refers to the hot 
component, which consists of diffuse ejecta and swept-up circumstellar matter.  
The cool clumpy is the sum of the cool X-ray and optical components also
assumed to be ejecta. The total is the sum of all the components.}
\centering
\begin{tabular}{|c|c|ccccc|}
\hline
 & & $v_{\rm r}$ & $M_{\sun}$ & $E_{\rm th}$ & $E_{\rm r}$ & $E_{tot}$  \\
 & &     km\,s$^{-1}$ & & $10^{43}$ J & $10^{43}$ J & $10^{43}$ J\\
\hline
1&hot CSM: & 1740 & 7.9 & 6.5 & 2.4 & 8.9 \\
2&hot ejecta     & 1740 & 0.4 & 0.3 & 0.1 & 0.4 \\
3&warm ejecta    & 1780 & 1.7 & 0.2 & 0.5 & 0.7 \\
4&cold ejecta    & 5290 & 0.1 & 0.0 & 0.3 & 0.3 \\
 \hline
\hline
$\Sigma_{1}^{2}$    &hot diffuse& -- & 8.3 & 6.8 & 2.5 & 9.3 \\
$\Sigma_{3}^{4}$    &cool clumpy& -- & 1.8 & 0.2 & 0.8 & 1.0 \\
$\Sigma_{2}^{4}$  &ejecta     & -- & 2.2 & 0.5 & 0.9 & 1.5 \\
\hline
$\Sigma_{1}^{4}$&total      & -- & 10.1& 7.0 & 3.3 & 10.3 \\
\hline
\end{tabular}
\label{tab4}
\end{table}
The top half of Table is the remnant as we see it now and summing up
all the energy components gives an estimate of the total energy released
by the explosion, $10.3\times10^{43}$ J. The cool X-ray
component, the optical knots and the heavy elements in the hot X-ray
component are almost certainly all remains of the ejecta. The evidence for 
the X-ray components being emission from ejecta material was put put forward by 
Willingale et al. (2002).
This evidence includes: i) the elemental abundances of 
Si, S, Ar and Ca are strongly correlated, ii) the elemental abundance 
values are consistent with enrichment from ejecta material iii) the emitting 
material is non-uniformally distributed across the remnant. In particular the 
emission of Fe-K relative to the lighter elements indicate a large degree of 
non-uniform mixing. The mass labelled hot ejecta in Table \ref{tab4} are the 
heavy elements (not H+He) of the hot component. The presence of these elements
in the hot component indicates considerable mixing between the swept
up interstellar material and the diffuse ejecta.
Previous authors, for example Anderson and Rudnick (1995),
have identified this component as diffuse ejecta and
estimated the mass as $\sim0.3$ $M_{\sun}$ (Braun 1987),
very similar to our estimate of 0.4 $M_{\sun}$. 
The remaining mass of 7.9 $M_{\sun}$ in the hot component is swept up material
labelled hot CSM in Table \ref{tab4}.

Summing up the mass from all the ejecta components indicated
in Table~\ref{tab4} and
assuming all the energy was kinetic we get the
predicted rms ejecta velocity of 6850 km s$^{-1}$.
The original kinetic energy from the 
diffuse (now hot) ejecta constitutes the driving mechanism of 
the primary blast wave. Assuming that all the energy in the hot diffuse medium 
(i.e. $9.3{\times}10^{43}$~J) was kinetic would require the hot ejecta 
(0.4~$M_{\sun}$) to have an initial velocity of
$\sim1.5{\times}10^{4}$~km\,s$^{-1}$.
We discuss this (quite feasible) velocity further in 
Sect. \ref{sect6}.
The original rms velocity of the clumpy (now cool/cold) ejecta
must have been much lower, $\sim2400$~km\,s$^{-1}$.
The velocity of the optical
FMFs is 8820 km\,s$^{-1}$ with a deceleration parameter
of 0.99 and the velocity of the optical FMKs is 5290 with a deceleration
parameter of 0.98
(see tabulation in Willingale et al. 2002 and references ibid) so
these ejecta velocity estimate are entirely reasonable. Estimation of
the mass and energy in two identifiable parts of the ejecta
gives us the first observational glimpse at the ejecta structure function 
which plays an important role in analytical and numerical modelling
of the early stages of the evolution of SNR, see Truelove and McKee (1999).
The values in Table \ref{tab4} are subject to uncertainties
which will only be resolved by observations which much higher spectral
resolution but
overall the mass and energy budget balances reasonably well.

The hot and cool components also contain 0.012 and 0.046 $M_{\sun}$
of iron respectively.
It is reasonable to assume that almost all of this iron
originated in the ejecta rather than from swept up interstellar medium
since most of
the material surrounding the star prior to the explosion probably came
from the outer hydrogen rich layers of the progentitor (see discussion below).
If this is the case $\sim2.7\%$ of the diffuse and clumpy
ejecta mass was iron, now seen as Fe K emission from the hot component
and as Fe L
emission from the cool component. The hot iron has a significantly larger
radial velocity, 2000 km\,s$^{-1}$, than the cool iron,
1580 km\,s$^{-1}$, and is seen at larger radii.
It is surprising that
iron is seen in the diffuse ejecta especially at large radii in the remnant
ahead of the lighter elements
since it presumably originated from
the core of the progenitor not the outer layers of the star.
A great deal of mixing of the layered structure of the progenitor
must have occured. This may be because the inner layers were ejected
at higher initial velocity than the outer layers and this, in turn,
resulted in significant turbulence.

\section{Discussion\label{sect6}}

Just how robust are the values in Table \ref{tab4}? Greatest uncertainty
lies in the measurement of the ion temperature and estimation of the
volume filling factors. We have set the maximum $\eta_{\rm hot}+\eta_{\rm 
cool}=1$
while the mean value is $\sim0.3$ which is entirely consistent with the observed
angular coverage shown in Fig. \ref{fig5}. If we abandon pressure
equilibrium the $\eta_{\rm cool}$ could increase but the $\eta_{\rm hot}$
would have to decrease and/or the overall filling factor would have to fall.
The mean ion temperature is constrained by the measured Doppler
broadening of the emission lines. The cool ions could be hotter raising the
cool pressure and introducing a pro rata increase in $\eta_{\rm cool}$.
This would increase the ejecta mass and decrease the swept up mass but
the total mass and energy would remain approximately the same. This, in turn,
would decrease the rms velocity of the ejecta which at present is 
consistent with the measured expansion velocities of the optical knots.

We have not included the magnetic pressure (or energy) in the calculations.
The electron pressure in the hot component is $4.3\times10^{-9}$ Pa,
only $\sim5\%$
of the total pressure and $7.6\times10^{-9}$ Pa, $\sim10\%$ of the
total pressure in the cool component.
A magnetic field of $\sim2.9$ mG will give the same
energy density as the electrons in the hot component but such a large
equipartition field is unlikely
since the field is being amplified by turbulence and magnetic coupling in
the post shock plasma. The mean magnetic field required assuming
equipartition with the high energy electrons responsible for the radio
synchrotron emission is $\sim0.5$ mG (Rosenberg 1970, Longair 1994).
We conclude that the magnetic energy and relativistic electron energy
are only a minor perturbation on the overall energy budget.

Very little of the mass and energy in Table \ref{tab4} is associated with
the faint primary shock which is visible in the Chandra X-ray image
(Hughes 2000) and radio images (Anderson \& Rudnick 1995).
Analysing the Chandra image we find only $\sim12\%$ of the X-ray flux lies
outside the main ring of emission in a region that
could be directly associated
with the primary shock. It may be that the mass and energy of the primary
shock are invisible because the electron temperature has not yet
reached the threshold required for X-ray emission. However, we think
this is unlikely since the modelling of Laming (2001) indicates that
the electrons should reach a temperature of $\sim3\times10^{7}$ K
only 10 years after being shocked and this translates to an angular shift
of 2.5\arcsec\ on the sky for a shock moving at 4000 km\,s$^{-1}$.
It is also possible
that slow ejecta lie inside the X-ray emitting shell and this 
material will remain
invisible until it is enveloped by the reverse shock. This could increase
the ejecta mass estimate
but would have only a minor effect on the total energy.
We conclude that
hidden mass or energy are unlikely to increase the budget in Table \ref{tab4}
by more than a few percent.

The swept up mass is only seen over about
40\% of the total volume of the remnant, Fig. \ref{fig5}.
Using an outer radius of 160\arcsec\ the implied density of the ambient medium 
within this volume before the explosion
was $\sim13$ cm$^{-3}$ which is higher than previous estimates inferred from
H II emission (8 cm$^{-3}$, Peimbert 1971)
deceleration of the radio-emitting material (2 cm$^{-2}$, Braun 1987)
or the low temperature of the X-ray emitting ring (McKee 1974).
This high density CSM provides a link between Cas A and a Wolf-Rayet progenitor
that suffered mass-loss forming
a nebular shell prior to the supernova explosion. Nebular shells around 
Wolf-Rayet stars are well observed phenomena. They typically have radii 
of the order $R\sim 1-4$~pc with shell thickness ${\Delta}R\sim 0.01-0.2$~pc
and 
electron densities $n_{\rm e}$ of a few tens to a few hundreds cm$^{-3}$ 
(Esteban et al. 1993).
We suggest some fraction of 
a nebula shell was initially heated by the primary shock to form the
hot CSM entry in Table~\ref{tab4}. The hot ejecta have been mixed with
this material by turbulence.
From the difference between the assumed age of the 
remnant and the time since the the gas was shocked (see the ionisation age in 
Table~\ref{tab3}), we estimate that the system was in free expansion for the 
first $\sim$100 years. This time, coupled with the initial velocity 
($1.5{\times}10^{4}$~km\,s$^{-1}$) derived in Sect. \ref{sect5},
implies that the material 
that we now see as hot ejecta had travelled out to a distance of $\sim$1.5~pc 
(90\arcsec) before it hit, and heated, the putative dense
nebular shell shed by the progenitor WR star. It then sweeped up about 8 
$M_{\sun}$ and is now entering the Sedov phase.
These dimensions are consistent with both 
the observed size of the 
emitting shell and the typical size of nebular shells observed around WR stars. 
The cooler ejecta, which have a much lower initial velocity,
impacted later and show 
a broader distribution of ionisation ages centered around a lower 
average absolute value (see Fig.~\ref{fig3}). The emission we see from this 
component is ablated material formed by ``reverse'' shock-heating of cool, 
clumpy ejecta.

The total {\em visible} emitting mass calculated above is lower than
previous estimates, Fabian et al. (1980) 15 $M_{\sun}$, Vink et al. (1996)
14 $M_{\sun}$.
The difference is largely attributable to the lower volume estimates.
With better spatial resolution and the benefit of Doppler measurements
the estimate of the total fraction of the shell volume which is occupied by
the emitting plasma is considerably reduced. The spatially resolved
high resolution X-ray spectra provided by XMM-Newton also give us
a detailed inventory of the state and composition of the plasma
which also reduces the uncertainty in estimating the masses involved.
However, the mass loss from 
the progenitor could still be $\sim20$ $M_{\sun}$ indicative of
a very high loss rate prior to the explosion although only
$40\%$ of this material is actually visible.

\section{Concluding remarks}

We have assumed pressure equilibrium between the hot and cool plasma
components to give an estimate of the filling factors within the shell
volume. X-ray spectroscopy at higher spectral and 
spatial resolution could be used
to test this assumption. Well resolved emission lines from individual knots
would be associated with either the hot or cool component
and observation of
Doppler broadening of such lines would give us a direct
measurement of the ion velocities for individual ion species in the ejecta
and swept up material.

\begin{figure}[!htb]
\centering
\includegraphics[width=8cm]{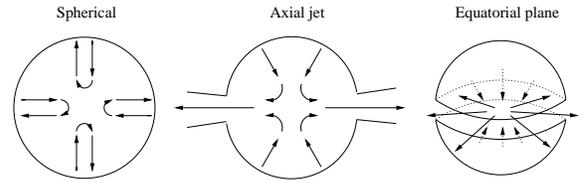}
\caption{Supernova collapse geometries}
\label{fig7}
\end{figure}

Possible supernova core collapse geometries are shown in Fig. \ref{fig7}.
The distribution of X-ray emitting mass around Cas A indicates that
the original explosion was not symmetric but somewhere between
an axial jet and equatorial plane geometry. The confinement to within
$\pm30$\degr of the
equatorial plane as shown in Fig. \ref{fig6} is rather striking
and the other panel in Fig. \ref{fig6}
clearly demonstrates the enhancement of the
emission around the poles in the axial coordinate system.
It is noteworthy that spherical collapse can be modelled in one dimension,
and the axial or equatorial symmetry can be modelled using just
two dimensions but the combination of axial and equatorial would require
a full three dimensional treatment. It may be that the processes responsible
for what we observe will only be revealed by such three dimensional modelling.
The apparent asymmetry of the explosion
geometry introduces the possiblity of significant shear within the
expanding material during or just after the explosion.
This may be the root cause of the turbulence and
the clumpiness of the mass distribution in the remnant rather than
hydrodynamic instabilities in the dense shell formed much later after
a significant mass of surrounding material has been swept up.

The total kinetic energy derived for the ejecta is consistent with the canonical
value of $10^{44}$ J. However measurements suggest that the ejected
mass was rather large and the rms ejection velocity was correspondingly
modest. Cas A was most emphatically
a mass dominated rather than radiation dominated supernova explosion.
This is in stark contrast with, for example, the Crab Nebula in which
no significant ejected mass or energy from the original explosion
has been identified, see for example Hester et al. (1995).
Collimated or jet-induced hypernovae have been suggested as a possible
solution to the energy budget problem posed by gamma ray bursts seen
from cosmological distances, Wang \& Wheeler (1998). However in these
cases we are looking for collimation in a radiation dominated
explosion. There is no reason to suppose that the degree of mass collimation
seen in Cas A is connected with radiation collimation inferred in gamma
ray burst events.

\begin{acknowledgements}
The results presented are based on observations obtained with XMM-Newton,
an ESA science
mission with instruments and contributions directly funded by
ESA Member States and the USA.
\end{acknowledgements}


\begin{thebibliography}{}
\bibitem[1995]{anderson} Anderson, M.C., \& Rudnick, L., 1995,
ApJ, 441, 307
\bibitem[1995]{aschenbach} Aschenbach B., Egger, R. \&
Tr\"{u}mper, J. 1995, Nature, 373, 587
\bibitem[1987]{braun1} Braun, R. 1987, A\&A, 171, 233
\bibitem[1987]{braun2} Braun, R., Gull, S.F., \& Perley, R.A. 1987,
Nature, 327, 395
\bibitem[1993]{esteban} Esteban, C., Smith L.J., V\'{i}lchez, J.M., \& Clegg, 
R.E.S., 1993, A\&A, 272, 299
\bibitem[1980]{fabian} Fabian, A.C., Willingale R., Pye J.P., Murray S.S.,
Fabbiano G. 1980, MNRAS, 193, 175
\bibitem{hst}Hester, J.J., Scowen, P.A., Sankrit, R., et al.
1995, ApJ 448, 240
\bibitem[2000]{hughes} Hughes, J.P., Rakowski, C.E., Burrows, D.N.,
\& Slane, P.O. 2000, ApJ, 528, L109
\bibitem[2001]{laming} Laming, J.M. 2001, ApJ, 563, 828
\bibitem[1994]{longair} Longair, M.S. 1994, High Energy Astrophysics,
Vol.2, Cambridge Univ. Press
\bibitem[1983]{markert} Markert, T.H., Canizares, C.R., Clark, G.W.,
\& Winkler, P.F., 1983, ApJ, 268, 13
\bibitem[1974]{mckee} McKee C.F. 1974, ApJ, 188, 335
\bibitem[1995]{reed} Reed, J.E., Hester, J.J., Fabian, A.C.,
\& Winkler, P.F. 1995, ApJ, 440, 706
\bibitem[1970]{rosenberg} Rosenberg, I., 1970, MNRAS, 151, 109
\bibitem[1970]{truelove} Truelove, K.J. and McKee, C.F. 1999, Ap.J.Supp.Ser.
120, 299-326
\bibitem[1996]{vink} Vink, J., Kaastra, J.S., \& Bleeker, J.A.M. 1996, A\&A,
307, L41
\bibitem[1998]{wang} Wang, L. \& Wheeler, J.C., 1998, ApJ, 504, L87
\bibitem[2002]{willingale} Willingale R., Bleeker J.A.M., van der Heyden K.J.,
Kaastra J.S., \& Vink J., 2002, A\&A 381, 1039-1048
\end{thebibliography}
\end{document}